\documentclass{iau} 
\usepackage{graphicx}

\title[Similarity and diversity of black holes - view from the Very High Energies ] 
{Similarity and diversity of black holes - view from the Very High Energies}

\author[Elina Lindfors]   
{Elina Lindfors$^1$}

\affiliation{$^1$Tuorla Observatory, Department of Physics and Astronomy, University of Turku, Finland \\ email: {\tt elilin@utu.fi}}

\pubyear{2017}
\volume{324}  
\jname{New Frontiers in Black Hole Astrophysics}
\editors{A.C. Editor, B.D. Editor \& C.E. Editor, eds.}
\begin{document}

\maketitle

\begin{abstract}
Active galactic nuclei, hosting supermassive black holes and launching
relativistic jets, are the most numerous objects on the gamma-ray sky.
At the other end of the mass scale, phenomena related to stellar mass
black holes, in particular gamma-ray bursts and microquasars, are also
seen on the gamma-ray sky. While all of them are thought to launch
relativistic jets, the diversity even within each of these classes is
enormous. In this review, I will discuss recent very high energy gamma-ray results that
underline both the similarity of the black hole systems,
as well as their diversity. 
\end{abstract}

\firstsection 
\section{Introduction}

The known black hole systems cover both stellar mass and supermassive
black holes. In Active Galactic Nuclei, the activity is driven by the
accretion of matter to supermassive black hole in the centre of the
galaxy. In inactive galaxies the black hole may occassionally shine up due to tidal disruption events, which 
occur when a star gets too close
to the supermassive black hole and is bulled apart by the black hole's
tidal forces. In extreme and rare cases (only two have been observed \cite{cenko,komossa}) this launches a collimated jet
of particles. On stellar black hole mass scales, the microquasar
phenomenon consists of a black hole feeding from companion star and
launching a jet. Finally, we assume that long gamma-ray bursts are
death crowls of massive stars, creating a black hole and collimated
jets of particles.

The common astrophysical ingredients of these systems, the spinning
black hole, the accretion disk, and the collimated jets of particles
have led several scientists to look for similarities among the
different systems. 
What properties simply scale with the black hole mass and could this be
signature of something fundamental? For example, in the early work by
\cite{merloni} a fundamental plane of black hole activity was
established, when the authors found that active galaxies and galactic
black holes lie on a plane in three dimensional space (radio
luminosity, X-ray luminosity and mass of the black hole). More
recently, \cite{nemmen} found that jets created by black holes maintain
the same coupling between the total power carried by the jet and power
radiated away. In our work, we have investigated if the jets of
blazars and microquasars are similar in terms of outbursting mechanism
and jet parameters that can be derived from the decomposition of the
radio to optical light curves (\cite{turler99}). We found that indeed
the outburst of the microquasars and quasars are well described by
shock-in-jet model, and that the jet parameters derived were rather
similar for both types of systems (\cite{turler07}).

In recent years, astroparticle physics has opened new observational
window to black hole systems and during our symposium we heard several
interesting presentations coming from this community. Very High Energy (VHE, $E>100\,GeV$)
$\gamma$-ray experiments have revealed $>170$ astrophysical sources,
and black hole systems are well represented among these
sources. IceCube has started the era of neutrino astronomy with the
discovery of neutrinos of astrophysical origin (\cite{IceCube}) and
while their origin is still unknown, black hole systems are certainly among
the candidates. Finally, the LIGO experiment measured gravitational
waves from merging black holes (\cite{ligo}). In this paper, I
concentrate on the new observations of black hole system in the very
high energies, covering the observations of active galactic nuclei,
tidal disruption events, microquasars and gamma-ray bursts.

\begin{figure}[b]
\begin{center}
 \includegraphics[width=4.0in]{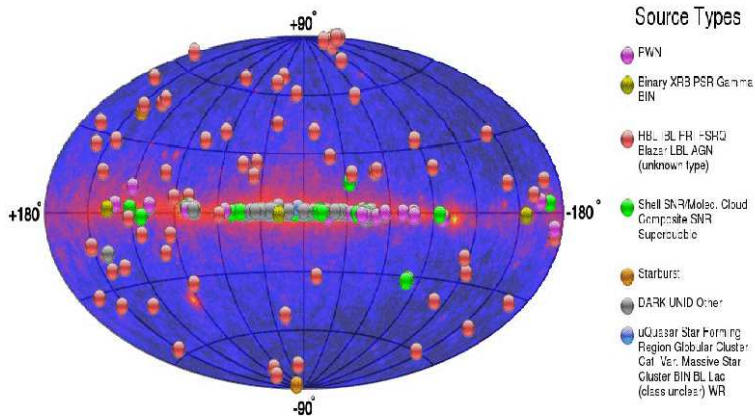} 
 \caption{The Very High Energy gamma-ray sky. The map from http://tevcat.uchicago.edu.}
   \label{fig1}
\end{center}
\end{figure}

\section{Supermassive black holes at very high energies}

\subsection{Active Galactic Nuclei}

\begin{figure}[b]
\begin{center}
 \includegraphics[width=4.0in]{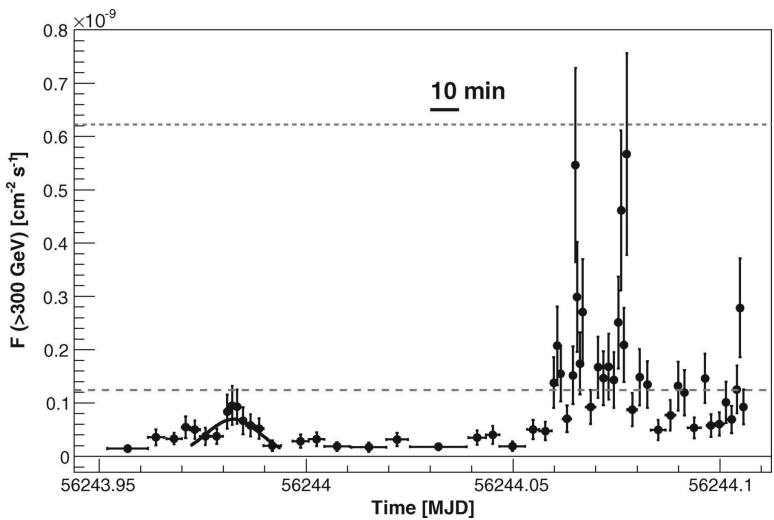} 
 \caption{The light curve of IC~310 on the night of 12/13 November 2012 as observed with the MAGIC telescopes. Figure from \cite{ic310}.}
   \label{fig1}
\end{center}
\end{figure}

Active galactic nuclei (AGN) are the most numerous sources on the
extragalactic VHE $\gamma$-ray sky. However, there is quite some
diversity among this population. The most numerous ones are blazars, a
type of active galactic nuclei, where the relativistic jet points very
close to our line of sight. The blazars divide into several
subcategories, all of which are seen at the VHE $\gamma$-ray
sky. Furthermore, also several radio galaxies are detected.

Blazar spectral energy distribution (SED) shows two bumps, the low
energy bump extends from radio to ultraviolet--X-rays while the high
energy bump extends from X-rays to VHE $\gamma$-rays. The low energy
emission is synchrotron emission by the electrons spiralling in the
magnetic field of the jet, while the high energy emission is in most
cases inverse Compton emission. The location of the synchrotron peak
is used to divide the blazars in sub-categories. The sources having
the peak at UV-X-ray energies are the most numerous sources in the
extragalactic VHE $\gamma$-ray sky, but nowdays also low synchrotron
peaking objects have been detected.

Blazars are variable in all wavelengths from radio to VHE
$\gamma$-rays in timescales ranging from minutes to years. While
short timescale variability is expected, as the jet is pointing so
close to our line of sight, it becomes challenging when the timescales
are as short as minutes. However, current instruments have observed
minute-scale variability in VHE $\gamma$-rays from several sources,
from all blazar classes and also from a radio galaxy
(\cite{2155,mrk501,pks1222,bllac,ic310}).

The minute-scale variability is challenging for models, as a huge
energy has to be radiated within an extremely compact region. The time
scale is actually shorter than the scale expected from the central
black hole's horizon (which for $10^{9}M_{sun}$ black hole is an order
of one hour). Furthermore, it is a challenge for particle acceleration
and emission models. Many solutions to this dilemma has been
suggested, such as strong recollimation of the jet or very compact
region embedded within large scale jets (spine-sheath, jets-in-jet,
ring-of-fire) (e.g. \cite{ghisellini05,giannios,macdonald}).

In many cases the fast VHE $\gamma$-ray variability is detected during
periods when the source has been showing enhanced flux levels in all
wavebands already prior to the detection of the fast variability. It
is therefore evident that these events somehow connect to the overall
activity within the relativistic jet. In the particular case of flat
spectrum radio quasars, it is also evident that the VHE $\gamma$-ray
emission cannot originate very close to the central black hole, as it
is surrounded by broad line emission clouds (at the distance of
$\sim$1 parsec) that are very efficient in absorbing VHE $\gamma$-rays
via pair production (e.g. \cite{bottcher16} and references therein).

One particularly interesting case of fast variability is radio
galaxy IC 310 in Perseus cluster. MAGIC Telescopes detected a huge
flare in the VHE $\gamma$-ray band in November 2013. The flux reached
several Crab Units, and the light curve revealed variability with
doubling timescales faster than 4.8 minutes (\cite{ic310}). As the
viewing angle of the jet is rather well constrained to $10-20$
degrees, the general solution of introducing $\Gamma>50$ to explain
the fast variability does not work. In general it was concluded that
models, where such a bright flare with such fast variability would be
produced within the jet, were not viable. Therefore it was suggested
that the flare actually originated in the magnetosphere of the black
hole and it was dubbed a black hole lightning. The magnetospeheric
model for IC 310 was discussed in detail in \cite{hirotani}, who
concluded that it is feasible to produce the observed flare if the
black hole at the time of the flare was accreting at very low
rate. The feasibility of this model was discussed also during the
symposium (see Barkov, this volume).

Beyond the fast variability of VHE $\gamma$-rays, also the slower
timescale variability of the VHE emission, the shape of the VHE
$\gamma$-ray spectrum and in more general terms the population of
active galaxies that we see at the highest energies tell us important
stories about supermassive black holes. Looking at the variability of
the light curves in different bands as well as the snap-shot spectral
energy distributions have shown that the emission takes place in
multiple emission regions within the relativistic jets
(e.g. \cite{1424,mrk421,Lindfors}). It has also been shown that, at least
occasionally, the main energy dissipation region must move further out
in the jet, at least to distances beyond the broad line region clouds
(\cite{ghisellini13,pks1510,pks1441}).

\subsection{Tidal Disruption Events}

\begin{figure}[b]
\begin{center}
 \includegraphics[width=3.9in]{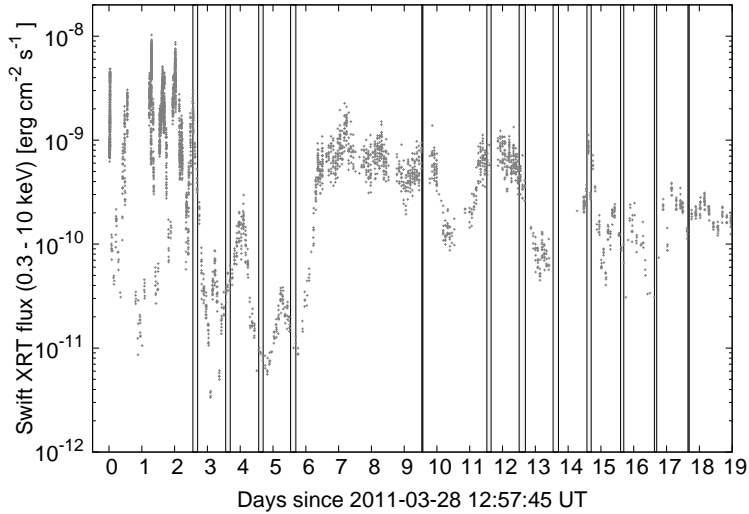} 
 \caption{The light curve of Sw 1644+57 from \textit{Swift}-XRT. The MAGIC observation windows of the event are shown as vertical lines. MAGIC detected no VHE $\gamma$-rays from this TDE. Figure from \cite{tde}.}
   \label{fig2}
\end{center}
\end{figure}

In addition to AGN, the VHE $\gamma$-ray
telescopes have been pointed to a normally inactive supermassive black hole, to the famous tidal disruption event Sw 1644+57.
\cite{bloom} and \cite{burrows} argued that this event arose from the
activation of a beamed jet that was hypothesized to be the result of a
tidal disruption of a star by a $\sim10^{6}-10^{7}M_{\odot}$ black hole.

The VERITAS Telescopes started observing Sw J1644+57 approximately
22.5 hours after the first BAT trigger and followed it for 18 days
(\cite{veritas_tde}). The MAGIC telescopes observed the Sw J1644+57
during the flaring phase, starting observations nearly 2.5 days after
the trigger time (\cite{tde}). Neither of the telescopes found evidence
for emission above the energy threshold of 100 GeV and the upper limits
were in agreement both with the synchrotron and inverse Compton
scattering scenarios for the X-ray emission (\cite{veritas_tde,tde}).
However, in principle the VHE $\gamma$-ray observations can be very
constraining for the conditions of the jet in such tidal disruption
events, in particular for the synchrotron origin of the X-ray emission
and for the lorentz factor of the newly formed jet.

In addition to Sw 1644+57, only one other possible case of jetted
tidal disruption event is known (\cite{cenko,komossa}), but to my
knowledge it was not followed by the VHE $\gamma$-ray telescopes.

\subsection{Supermassive black hole in the centre of our Galaxy}

Galactic center is a strong and well established source of VHE
$\gamma$-rays (\cite{hess_gc,magic_gc,veritas_gc}). It was already
early suggested that the point source in the galactic centre would
correspond to a central black hole (\cite{hess_gc}), but as there are also
other candidates, such as supernova remnant Sgr A East, the
question of direct emission of the black hole in the center of our
galaxy is not yet resolved.

Very recently H.E.S.S. Collaboration showed, using 10 years of VHE
$\gamma$-ray observations of the galactic centre, that the black hole
in the centre of our galaxy is the first established PeVatron, i.e
particle accelerator that can accelerate particles up to PeV energies
(\cite{hess_pev}). 

\section{Stellar mass black holes at very high energies}

\subsection{Microquasars}
X-ray binaries are binary systems with a compact object (a black hole
or a neutron star) feeding from a companion. In microquasars this
feeding launches a relativistic jet and as an analogy to more massive
quasars, these jets could accelarate particles to energies high enough
to emit VHE $\gamma$-rays. 

There were two detections of VHE $\gamma$-ray emission from binary
systems that were first considered as possible microquasars, LS5039
and LSI +61 303 (\cite{ls5039,lsi}), but later observations have
supported binary pulsar model for these sources
(e.g. \cite{magic_lsi08}). Up to date, the hint of VHE $\gamma$-ray
emission from a microquasar Cyg X-1 (\cite{cygx1}) is the only
indication of VHE emission, while other observations have resulted only in
upper limits (\cite{magic_cygx3,hess_ul}). The reason can be that the
conditions are not favorable to emission of VHE $\gamma$-rays, given
that it always requires acceleration of particles to extreme energies
and typically also presence of photons to be up-scattered. Other
possible explanation is an observational bias, as it is evident
that the high energy emission from the microquasar jets is a transient
phenomena.

Cyg~X-1 and Cyg~X-3 are both detected in the lower gamma-ray
energies by the Fermi satellite and the origin of this emission seems
to be the jet (\cite{fermi_cygx3,zanin,zdziarski}). Therefore, it is to be
expected that during flares the emission would extend also to energies
$>100$ GeV, unless the gamma-rays are always emitted very close to jet
base, where strong photon field absorbing VHE $\gamma$-rays is
present. If the microqusars were to be analogies to quasars, one would
expect that occassionally the energy dissipation region would move
further out and VHE $\gamma$-rays could escape. To catch such a epoch,
which is assumably an order of an hour in duration or shorter, with ground-based
telescopes (limited duty cycle and limited field of view) is of course
challenging. 

\begin{figure}
\begin{center}
 \includegraphics[width=4.0in]{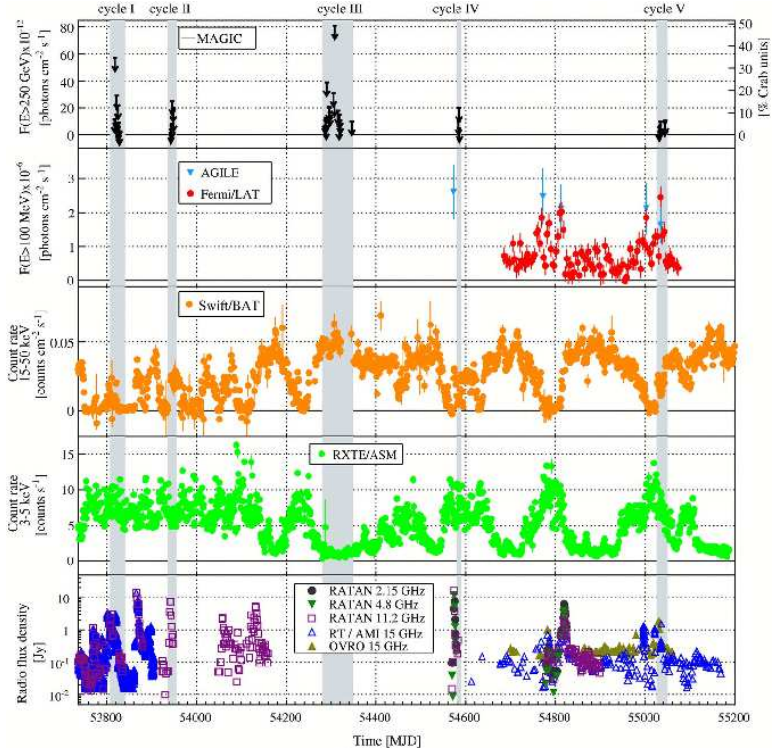} 
 \caption{The light curve of Cyg~X-3 as observed with the MAGIC telescopes (top), textit{Fermi}-LAT and AGILE satellites, \textit{Swift}-BAT, RXTE-ASM and OVRO and AMI telescopes (bottom) from 2006 to 2009. The gray band show the periods corresponding to MAGIC observations. The MAGIC observations covered different X-ray/radio spectral states. No VHE $\gamma$-ray emission from the source was detected. Figure from \cite{magic_cygx3}.}
   \label{fig2}
\end{center}
\end{figure}

\subsection{Gamma-ray bursts}

Gamma-ray bursts (GRBs) launch short lived, but extremely fast and
luminous jets. As the duration of the prompt emission from GRBs is
typically an order of a minute or less, fast pointing to the location of
the GRBs is in a key role. Up to now there is no detection of VHE
$\gamma$-ray emission from GRBs by the Imaging Air Cherenkov
Telescopes (\cite{albert06,tam06,albert07,aharonian09,aleksic10, acciari11,aleksic14}), but the detection of high energy photons by the {\it Fermi}-LAT from some GRBs (e.g. \cite{lat_grb}) certainly gives hope also to ground-based experiments. VHE $\gamma$-ray observations are a powerful tool for emission processes and physical conditions in GRBs and could also give important insight on the similarity of the emission processes in GRB and blazar jets. 

\section{Summary}

Phenomena related to black holes at VHE $\gamma$-rays are mostly
transient or extremely variable and therefore a huge observational
challenge. As this window to black hole systems is still rather new,
there are many open questions and also significant discovery potential
in this energy range. The most prominent black holes at VHE
$\gamma$-rays are the active galactic nuclei, where we have already
started to see diversity as the number of known sources is steadily
increasing (currently $>70$). For the rest; tidal disruption events,
microquasars and gamma-ray bursts, we will have to wait for further
observations at VHE $\gamma$-rays, with the current generation of
telescopes and with the future Cherenkov Telescope Array, to be able
to say something about the similarity of the all black hole systems at
very high energies.

\end{document}